\documentclass[prl,article,preprint,groupedaddress,12pt]{revtex4}
\usepackage{epsfig}
\usepackage{graphicx}
\usepackage{amssymb}

\begin{document}
\title{A glimpse into the magical world of quantum gravity}
\author{Shahar Hod}
\affiliation{The Ruppin Academic Center, Emeq Hefer 40250, Israel}
\affiliation{ } \affiliation{The Hadassah Institute, Jerusalem
91010, Israel}
\date{\today}
\centerline {\it This essay received an Honorable Mention from the Gravity Research Foundation 2024.}

\begin{abstract}

\ \ \ In this essay it is proved that, in a self-consistent semiclassical theory of gravity, 
the asymptotically measured orbital periods of test particles around 
central compact objects are fundamentally bounded from below by the compact universal 
relation $T_{\infty}\geq{{2\pi e\hbar}\over{\sqrt{G}c^2 m^2_{e}}}$ [here $\{m_e,e\}$ 
are respectively the proper mass and the electric charge of the electron, the lightest charged particle]. 
The explicit dependence of the lower bound on the fundamental constants $\{G,c,\hbar\}$ 
of gravity, special relativity, and quantum theory suggests 
that it provides a rare glimpse into the yet unknown quantum theory of gravity.
\newline
\newline
Email: shaharhod@gmail.com
\end{abstract}
\bigskip
\maketitle

\subsection{Introduction}

While the basic physical principles of general relativity and quantum theory were already revealed about 
a century ago, we still lack a complete and self-consistent theory of quantum gravity. 
In particular, anyone who researches the interplay between quantum theory and gravitation knows 
how difficult it is to uncover the fundamental physical principles of the elusive microscopic 
quantum theory of gravity.

An important question that arises during the quest for a self-consistent quantum theory of gravity 
is how to identify basic physical principles which are expected to characterize this yet unknown 
fundamental theory of nature? 

In order to provide an answer to this physically interesting question, it is important to realize that any 
self-consistent combination of gravity with quantum theory is expected to yield new quantitative physical laws that 
are formulated in terms of the three fundamental constants of nature: The constant $G$ of gravity, the 
constant $c$ of special relativity, and the constant $\hbar$ of quantum physics \cite{Hol1,Hol2}.

In the present essay we shall reveal the existence of a fundamental physical principle which 
is expected to characterize any self-consistent semiclassical theory of gravity. 
In particular, we shall explicitly prove that, due to quantum effects in highly curved spacetimes, 
the asymptotically measured orbital periods of test particles around central compact objects 
are fundamentally bounded from below by a remarkably compact 
relation which is formulated in terms of the fundamental constants of nature: $G, c$, and $\hbar$. 

\bigskip

\subsection{Universal lower bound on orbital periods around central compact objects}

Our main goal is to reveal the existence, within the framework of a self-consistent 
quantum theory of gravity, of a fundamental 
lower bound on orbital periods of test particles around charged compact objects. 

The curved spacetime generated by a spherically symmetric charged compact object of an 
asymptotically measured mass $M$, electric charge $Q$ \cite{NoteQ0}, 
and radius $R$ is characterized by the line element \cite{Bar,Chan,Shap}
\begin{eqnarray}\label{Eq1}
ds^2=-f(r)dt^2+
f^{-1}(r)dr^2
+r^2d\theta^2+r^2\sin^2\theta d\phi^2\ \ \ \ \ \text{for}\ \ \ \ \ r\geq R\  ,
\end{eqnarray}
where the radially dependent metric function in (\ref{Eq1}) is given by the expression
\begin{equation}\label{Eq2}
f(r)=1-{{2M}\over{r}}+{{Q^2}\over{r^2}}\  .
\end{equation}
We shall use natural units in which $G=c=\hbar=1$ throughout most of the essay. 
However, in the summary section we shall emphasize the explicit dependence of our analytically derived 
lower bound [see Eq. (\ref{Eq13}) below] on the fundamental constants $\{G,c,\hbar\}$ of nature. 

Circular motions of test particles around central compact objects 
provide important information about the physical parameters (like mass and electric charge) that characterize 
the corresponding curved spacetimes \cite{Bar,Chan,Shap}. 
In particular, the asymptotically measured orbital period $T_{\infty}$ of a test particle 
around a central compact object is of fundamental importance in theoretical studies as well as in observational studies 
of curved spacetimes. 

In the present essay we shall address the following physically interesting question: 
Do the basic laws of a self-consistent quantum theory of gravity set a fundamental lower bound 
on the asymptotically measured orbital periods of test particles in curved spacetimes of highly 
compact objects?

In order to address this physically important question, we first note that an orbiting particle may 
use non-gravitational forces in order to move arbitrarily close to the speed of light 
along a non-geodesic circular trajectory. 
Thus, in order to determine the {\it shortest} possible 
orbital period around the central charged compact object (\ref{Eq1}), we shall consider highly relativistic 
test particles that move arbitrarily close to the speed of light. 

For such highly relativistic particles 
the orbital period $T_{\infty}$ as measured by far away (asymptotically flat) observers can be 
deduced from the line element (\ref{Eq1}) with the relations \cite{Notethet,Hodep}:
\begin{equation}\label{Eq3}
ds=dr=d\theta=0\ \ \ \ \  \text{and}\ \ \ \ \ \Delta\phi=\pm2\pi\  .
\end{equation}
In particular, substituting the orbital properties (\ref{Eq3}) into the 
curved line element (\ref{Eq1}), one obtains the compact functional expression  
\begin{equation}\label{Eq4}
T_{\infty}(M,Q,r)={{2\pi r}\over{\sqrt{1-{{2M}\over{r}}+{{Q^2}\over{r^2}}}}}\
\end{equation}
for the asymptotically measured orbital period 
around the central charged compact object. 

Inspection of Eq. (\ref{Eq4}) reveals the physically interesting fact that, for a given mass $M>0$ of 
the central compact object and for a given radius $r$ of the circular trajectory, 
the asymptotically measured orbital period $T_{\infty}(Q;M,r)$ is a monotonically decreasing function of 
the electric charge parameter $|Q|$ that characterizes the central compact object. 
The orbital period (\ref{Eq4}) can therefore be minimized by maximizing, for a 
given value $M$ of the total gravitational mass, the electric charge parameter of the compact object. 

In particular, one obtains from (\ref{Eq4}) the relation 
\begin{equation}\label{Eq5}
T^{\text{min}}_{\infty}(M,Q_{\text{max}},r)=
{{2\pi r}\over{\sqrt{1-{{2M}\over{r}}+{{Q^2_{\text{max}}}\over{r^2}}}}}\  ,
\end{equation}
where $Q_{\text{max}}=Q_{\text{max}}(r)$ represents the maximally allowed electric charge 
that can be contained within a sphere of radius $r$.

The compact mathematical expression (\ref{Eq5}) for the asymptotically measured 
orbital period around the central compact object 
naturally raises the following physically important question: 
Is there a fundamental physical mechanism which, within the framework of a self-consistent semiclassical theory of gravity, 
bounds the amount of electric charge that can be contained within a sphere of radius $r$? 

Interestingly, the answer to the above stated question is `yes'! 
\newline
In particular, the Schwinger mechanism of particle/anti-particle pair production by highly compact 
charged objects sets the fundamental quantum ($\hbar$-dependent) upper bound \cite{Sch1,Sch2}
\begin{equation}\label{Eq6}
{{Q(r)}\over{r^2}}\leq E_{\text{c}}\equiv {{m^2_e}\over{e\hbar}}\
\end{equation}
on the strength of the electric field which is generated by a charged compact object. 
Here the physical parameters $\{m_e,e\}$ are respectively the proper mass and the electric charge of the 
electron, the lightest charged particle in nature. 

Taking cognizance of Eqs. (\ref{Eq5}) and (\ref{Eq6}), one finds the compact expression
\begin{equation}\label{Eq7}
T^{\text{min}}_{\infty}(r;M,E_{\text{c}})={{2\pi r}\over{\sqrt{1-{{2M}\over{r}}+E^2_{\text{c}}r^2}}}\
\end{equation}
for the shortest possible orbital period, as measured by asymptotic observers, 
around a charged compact object of mass $M$.

Intriguingly, and most importantly for our analysis, inspection of the radius-dependent 
orbital period (\ref{Eq7}) reveals the fact that it can be {\it minimized} by 
the universal ($E_{\text{c}}$-independent) orbital radius
\begin{equation}\label{Eq8}
r^{\text{min}}=3M\  .
\end{equation}
In particular, substituting Eq. (\ref{Eq8}) into Eq. (\ref{Eq7}), one obtains the functional expression
\begin{equation}\label{Eq9}
T^{\text{min}}_{\infty}(M,E_{\text{c}})=2\pi\cdot\sqrt{{27M^2}\over{1+27M^2E^2_{\text{c}}}}\
\end{equation}
for the shortest possible orbital period, as measured by asymptotic observers, 
around a central charged compact object of total (asymptotically measured) mass $M$. 

The analytically derived expression (\ref{Eq9}) yields, in the dimensionless large-mass regime [see Eq. (\ref{Eq6})]
\begin{equation}\label{Eq10}
ME_{\text{c}}\gg1\ \ \Longleftrightarrow\ \  {{Mm^2_e}\over{e\hbar}}\gg1\  ,
\end{equation}
the remarkably compact semiclassical gravity bound
\begin{equation}\label{Eq11}
T^{\text{min}}_{\infty}(M,E_{\text{c}})\ \longrightarrow\ {{2\pi}\over{E_{\text{c}}}}\
\end{equation}
on orbital periods. 
Intriguingly, one finds that the the lower bound (\ref{Eq11}) is {\it universal} 
in the sense that it is independent of the asymptotically measured mass $M$ of the curved 
spacetime. 

\bigskip

\subsection{Summary and discussion}

In this essay we have raised the following physically interesting question: 
Do the basic laws of nature imply the existence of a fundamental lower bound on the orbital periods of test particles 
around compact objects?

In order to address this question we have analyzed the circular motions of highly relativistic 
test particles in curved spacetimes of charged compact objects. 
Intriguingly, we have revealed the physically important fact that, within the framework of a self-consistent 
semiclassical theory of gravity, the Schwinger mechanism \cite{Sch1,Sch2} 
of particle/anti-particle pair production by highly compact charged objects is responsible for the existence of 
the previously unknown fundamental lower bound [see Eqs. (\ref{Eq6}) and (\ref{Eq9})]
\begin{equation}\label{Eq12}
T_{\infty}(M)\geq T^{\text{min}}_{\infty}(M)=
2\pi\cdot\sqrt{{27M^2}\over{1+27M^2\cdot({{m^2_e}/{e\hbar}})^2}}\
\end{equation}
on asymptotically measured orbital periods. 

What we find most interesting is the fact that the analytically derived 
lower bound (\ref{Eq12}) becomes universal ({\it independent} of the asymptotically measured 
mass $M$ of the curved spacetime) 
in the dimensionless large-mass regime (\ref{Eq10}). 
In particular, written out fully in terms of the fundamental constants $\{G,c,\hbar\}$ of nature, 
the lower bound on orbital periods around central 
compact objects can be expressed in the remarkably compact form 
[see Eqs. (\ref{Eq6}) and (\ref{Eq11})] \cite{Notekk}
\begin{equation}\label{Eq13}
T_{\infty}\geq T^{\text{min}}_{\infty}={{2\pi\hbar\sqrt{k}e}\over{\sqrt{G}c^2 m^2_{e}}}\  .
\end{equation}

The intriguing fact that the fundamental constants $\{G,c,\hbar\}$ of gravity, special relativity, and quantum physics 
all appear in (\ref{Eq13}) 
suggests that this analytically derived relation is fundamentally important for a self-consistent formulation 
of the microscopic quantum theory of gravity. 

\newpage
\noindent
{\bf ACKNOWLEDGMENTS}

This research is supported by the Carmel Science Foundation. I would
like to thank Yael Oren, Arbel M. Ongo, Ayelet B. Lata, and Alona B.
Tea for helpful discussions.


\begin{thebibliography}{99}
\bibitem{NoteQ0} One may assume, without loss of generality, the relation $Q\geq0$ for 
the electric charge parameter of the central compact object.
\bibitem{Notethet} We shall consider, without loss of generality, equatorial trajectories ($\theta=\pi/2$) in the 
curved spacetime (\ref{Eq1}).
\bibitem{Notekk} Here $k=1/4\pi\epsilon_0$, where $\epsilon_0$ is the vacuum permittivity.
\end{thebibliography}
\end{document}